\documentstyle[11pt]{article}
\input epsf

\title{\bf Jet production in
pA and AA collisions in the perturbative QCD
pomeron model}
\author{M.A.Braun\\
Dep. of High Energy physics,
 University of S.Petersburg,\\
198504 S.Petersburg, Russia}
\def\beq{\begin{equation}}
\def\eeq{\end{equation}}
\def\noi{\noindent}
\begin{document}
\maketitle
\medskip
\noi{\bf Abstract.} Inclusive cross-sections for gluon jet
production are studied numerically in the perturbative QCD pomeron
model for pA and central AA collisions at high energies.
Two forms for the inclusive cross-sections, with and without
emission from the triple pomeron vertex, are compared. The
difference was found to reduce to a numerical factor $\sim 0.7\div 0.8$
for momenta below the saturation momentum $Q_s$. Above $Q_s$ no
difference was found at all. For pA collisions
the gluon spectrum was found to be
$\sim A^{0.7}$ at momenta $k$ below $Q_s$ and $\sim A^{0.9}$ above it.
For central AA collisions it was found to be
$\sim A$ at momenta $k$ below $Q_s$ and $\sim A^{1.1}$ above it. At
large  $k$ the spectrum goes like $1/k^{2.7\div 3.3}$ flattening
with energy. The multiplicities turned out to be proportional to
$A^{0.7}$ for pA collisions and $A$ for central AA collisions
with a good precision. In the latter case they are becoming more
peaked at the center with the growth of energy.
Their absolute values are high and grow
rapidly with energy in accordance with the high value of the BFKL
intercept. \vspace{1.5cm}

\section{Introduction}
New experimental data on heavy-ion collisions  at RHIC and
future such data to be obtained at LHC attracted much attention to
the spectra of produced secondaries and in particular their
dependence on the atomic number of colliding nuclei. One would like
to have relevant
predictions  based on the
fundamental theory and not purely phenomenological. At present
the only candidate for this is the hard pomeron model derived
from perturbative QCD. Originally constructed for the description
of high-energy low -$x$ hadronic scattering  (the BFKL model~\cite{BFKL})
it has subsequently been generalized to hadronic or deep inelastic scattering
on nuclei ~\cite{kov,bra1} and nucleus-nucleus scattering ~\cite{bra2}.
The model suffers from a
serious drawback related to the use of a fixed and not running
strong coupling constant. Curing it does not look too promising,
since due to absence of ordering of momenta in the model, it also
means solving the confinement problem. However in spite of this
defect the model seems to describe high-energy phenomena
in a qualitatively reasonable manner. Also  attempts to include
the running of the coupling in some effective way have shown that the
effect of the running is not at all overwhelming, although introduces
some quantitative changes into the predictions. So, also for lack of something
better, the perturbative QCD pomeron model appears to give a reasonable
basis for the discussion of particle production in very high-energy heavy-ion
collisions. Of course due to the  perturbative character of the model it
can only give predictions for production of jets, leaving jet-to-hadrons conversion
to non-perturbative fragmentation mechanism. Also it has to be stressed
that the model is, by construction, oriented towards very high
(asymptotic) energies
or, equivalently, very low values of $x$. So its application to
present-day energies does not seem to be fully justified, at least in the
lowest order of perturbation expansion to be used in the present
calculations. This has to be kept in mind when comparing the prediction
of the model with the existing experimental data.

From the start it has to be noted that in the model
scattering amplitudes can
be found comparatively easily only for  hadron-nucleus collisions.
They are given by a set of all   pomeron fan diagrams, which
are summed
by  the non-linear evolution equation
of ~\cite{kov,bra1}.
Nucleus-nucleus collision amplitudes
are described in the model  by complicated equations, whose solution is
quite difficult to obtain even numerically (see ~\cite{bra3}
for partial results).
Happily, as was shown in ~\cite{bra4}, due to Abramovsky-Gribov-Kancheli
(AGK)
cancellations  ~\cite{AGK}, to find the single inclusive distributions
one does
not have to
solve these equations, but only to sum the appropriate sets of fan diagrams,
that is to solve the much simpler hadron-nucleus problem.
Still the latter problem involves a numerical study of considerable
complexity.
So up to now there has been no consistent calculation of the jet spectra
for realistic nuclei at very high energies, although some preliminary
attempts has been done
in ~\cite{nestor,kov2, KNV,KLN}. In all cases however the authors relied
on very drastic
simplifications,
from the start choosing for the nuclear structure and/or for the gluon
distributions in the colliding  nuclei some primitive explicit forms
 in accordance with their own taste and prejudice. In fact these
forms appear to be  rather far from realistic ones, which correspond to
actual participants and follow from the calculations.
This gave us motivation to calculate numerically the jet spectra in
hadron-nucleus and nucleus-nucleus
collisions as predicted by the hard pomeron model in a consistent manner.

Another goal of the present calculations has been to compare the
results obtained on the basis of the expression for the inclusive
cross-section which follows from the AGK rules applied to the
diagrams with QCD pomerons interacting via the three-pomeron
coupling ~\cite{bra4} with a somewhat different expression obtained
from the colour dipole picture ~\cite{KT}. Our calculations show
that these two formally different expressions lead to completely
identical results at momenta of the order or higher than the value
of the so-called saturation momentum $Q_s$. At momenta substantially
lower than $Q_s$ the colour dipole cross-sections differ from the
ones from the AGK rules by a universal constant factor $\sim 0.7\div
0.8$.

In both cases the  $A$-dependence of the spectra at
momenta below  $Q_s$ is found to be
suppressed as compared to the naive probabilistic expectations,
which predict that it should follow the number of binary collisions,
$\propto A$ for pA scatering and $\propto A^{4/3}$ for
AA scattering at fixed impact parameter.
Insted we have found a behaviour damped by roughly factor $A^{1/3}$, that is
$\propto A^{0.7}$ for pA collisions and $\propto A$ for central
AA collisions.
Since $Q_s$
grows with energy very fast, the region where the spectra
behave in this manner
extends with energy to include all momenta of interest. At momenta
greater than $Q_s$ the spectra grow faster but still much slowlier
than the number of binary collisions
(a numerical fit gives something like $\propto A^{1.1}$ for central
AA collisions).

Note that in the last years a few more phenomenologically oriented
studies of particle production in nucleus-nucleus production at
very high energies have
been presented, in the framework of the color-condensate model
~\cite{ML} solved in the classical approximation on the lattice
~\cite{KNV} and in the saturation model ~\cite{KLN}.
 In both
approaches quantum evolution of the nuclear gluon density was neglected
and the saturation momentum was introduced as a parameter fitted to
the experimental data at RHIC. Although
 some of their predictions (proportionality
of the multiplicity to $A$ modulo logarithms) agree with our calculations
with full quantum evolutions, the quantitative results are rather different.
We postpone a more detailed discussion of this point until our Conclusions.

The results of this study continue and extend the ones
published in ~\cite{bra5}. As compared to that publication we
explore the $y$-dependence of the cross-sections and,
apart from AA central collisions, study pA collisons as well.
We also concentrate on energies accessible in the near
future (at LHC) to be closer to the experimental situation.
Still
we have to stress  that our results do not pretend to describe
the existing data obtained at RHIC. As mentioned, the perturbative QCD
pomeron
model studies phenomena at very high energies, at which the
coupling constant becomes sufficiently small.
The characteristic energetic variable for the model is the scaled
rapidity $\bar{Y}=(N_c\alpha_s/\pi)Y$.  One observes that the
solution of the non-linear evolution equation acquires its standard
scaling form, independent of the choice of the initial conditions,
at $\bar{Y}>1.5\div 2$. With $\alpha_s\sim 0.2$ this sets the lower
limit for the rapidity at the center, at which our results can be
applied, to be 7.5$\div$ 10, which implies the overall rapidity for
the collision in the region 15$\div$ 20. This is much higher than
the rapidities at RHIC but well within the possibilities of LHC.
So we may hope that our predictions can be tested at LHC but have
little chance of success for the data from RHIC.
In relation to
the latter  the model can possibly indicate some trends of
the observable cross-sections with the growth of energy, much in
the same manner as it predicted the growth of the total cross-sections
with energy at the time of its appearance (although the rate of this
growth was grossly overestimated).
The found proportionality of the spectra to the number of participants
and not to the number of binary collisions well agrees with the
experimental findings and is one of these trends.

\section{Basic equations}
\subsection{Total cross-sections}
As mentioned in the Introduction, the scattering amplitude can be
relatively easily found for hA collisions but not for AB collisions.
Corespondingly here we present formulas which serve to calculate
the total cross-sections for hadron (proton)-nucleus collisions.
This cross-section is given by an integral over the
impact parameter $b$ and pomeron transverse dimension $r$ as
\beq
\sigma_A(y)=2\int d^2b d^2r\rho(r)\Phi(y,r,b).
\eeq
Here $y$ is the overall rapidity, $\rho(r)$ is the colour dipole density
of the projectile (proton) and $\Phi(y,r,b)$ is the
colour dipole-nucleus cross-section at fixed $y,r$ and $b$, given by
the sum of all fan diagrams stretched between the projectile and nucleus.

Function $\phi(y,r,b)=\Phi(y,r,b)/(2\pi r^2)$,
in the momentum space,
satisfies the well-known
non-linear equation ~\cite{kov,bra1}
\beq
\frac{\partial\phi(y,q,b)}{\partial \bar{y}}=-
H\phi(y,q,b)-\phi^2 (y,q,b),
\eeq
where $\bar{y}=\bar{\alpha}y$, $\bar{\alpha}=\alpha_sN_c/\pi$,
$\alpha_s$ and $N_c$ are the strong coupling constant and the number
of colours,
respectively, and $H$ is the BFKL Hamiltonian. Eq. ( 2 ) has to be solved
with an initial condition at $y=0$ determined by the colour dipole
distribution in the proton smeared by the profile function of the
nucleus. In our calculations we take this distribution in accordance with
the Golec-Biernat distribution ~\cite{gobi}, duly generalized
for the nucleus:
\beq
\phi(0,q,b)=\frac{1}{2}AT_A(b)\sigma_0f(q),
\eeq
where
\beq
f(q)=-\frac{1}{2}{\rm Ei}(-x),\ \  x=\frac{q^2}{0.218\, {\rm GeV}^2},
\eeq
$\sigma_0=20.8$  mb
and $T_A(b)$ is the standard nuclear profile function, which we have
taken from the Woods-Saxon nuclear density.

To write the final formula for the cross-section we have to find the
dipole distribution $\rho(r)$ in the incoming proton, consistent
with the initial condition ( 3 ). To this end we use the expression for
the initial condition in terms of $\rho(r)$ (see ~\cite{bra1})
\beq
\Phi(0,r,b)=\frac{1}{2}AT_A(b)g_0^4\nabla^{-4}\rho(r),
\eeq
where $g_0$ is the strong coupling constant
at a very low  scale determined by the intrinsic momenta inside the
proton.  This translates into the relation
in the momentum space:
\beq
\rho(q)=-\frac{2\pi \sigma_0}{g_0^4}f_2(q),\ \
f_2(q)=\left(\frac{d}{d\ln q}\right)^2f(q).
\eeq
We also introduce the gluon density function
\beq
h(y,q,b)=q^2\nabla_q^2\phi(y,q,b)=\left(\frac{d}{d\ln q}\right)^2\phi(y,q,b)
\eeq
to finally obtain
\beq
\sigma_A(y)=\frac{\sigma_0}{8\pi^2\alpha_{s0}^2}\int d^2bd^2qf_2(q)h(y,q,b),
\eeq
where $\alpha_{s0}=g_0^2/(4\pi)$ refers to the scale inside the proton.
Of course  the sructure of the proton is unperturbative, so that
$\alpha_{s0}$ emerges as a new parameter in the model related to
hadronic processes.

Note that with these normalizations the cross-section at $y=0$ corresponds to
a pure double gluon exchange. A better approximation consists in
using an eikonalized anzatz for the initial condition:
\beq
\Phi(0,r,b)\to 1-e^{-\Phi(0,r,b)}.
\eeq
From the formal point of view this implies including higher orders in
$1/N_c^2$ and so cannot be justified. Also, as follows from calculations,
already at $\bar{y}=1$ the eikonalization becomes  forgotten
and the results with eikonalized and non-eikonalized initial conditions
are practically the same. Still at $y=0$ they are different and the
cross-section without eikonalization results unreasonably large.
For this reason we use the eikonalized initial condition for the
total hA cross-sections at low $y$ and the non-eikonalized one elsewhere.

\subsection{Inclusive cross-sections and multiplicities}
In the stydy of inclusive cross-sections we have to distinguish the
overall rapidity of the collision $Y$ and the rapidity of the produced
particle $y$.
Our basic quantity will be the inclusive cross-sections $I_{A}(y,k)$
and $I_{AB}(y,k)$
to produce a jet with the transverse momentum
$k$ at rapidity $y$ in a collision of a proton off a nucleus with the
atomic number $A$ or two nuclei with atomic numbers
$A$ and $B$:
\beq
I_{A(B)}(y,k)=\frac{(2\pi)^2d\sigma_{A(B)}}{dyd^2k}.
\eeq
Both cross-sections can be represented as an integral over the impact
parameter $b$:
\beq
I_{A(B)}(y,k)=\int d^2bI_{A(B)}(y,k,b).
\eeq
We shall study the pA cross-section as it stands, but for the nucleus-
nucleus case
our study will be restricted to the inclusive cross-sections
at a fixed impact parameter $b=0$ (central collisions) and identical
nuclei, $A=B$. To simplify notations we shall denote the AA cross-sections
also as $I_A$ in all places where it cannott lead to confusion.
The corresponding multiplicities at a fixed rapidity $y$
will be given by
\beq
\mu_{A}(y)=\frac{1}{\sigma_{A}}
\int\frac{d^2k}{(2\pi)^2}I_{A}(y,k)
\eeq
for the pA case and
\beq
\mu_{AA}(y)=\frac{1}{\sigma_{AA}(b=0)}
\int\frac{d^2k}{(2\pi)^2}I_{AA}(y,k,b=0),
\eeq
for the AA case,
where $\sigma_A$ and $\sigma_{AA}(b)$ are the total inelastic
cross-sections for the pA collision and for the
collision of two identical nuclei at the fixed impact parameter $b$.
For heavy nuclei one expects that $\sigma_A(b=0)\simeq 1$,
so that the multiplicity (13) is just the integral of the inclusive
cross-section over  momenta.

As argued in ~\cite{bra2}, in the perturbative QCD with a large number of
colours
the nucleus-nucleus interaction is described by a set of tree diagrams
constructed with BFKL pomeron Green functions and triple pomeron vertices
for their splitting and fusing. The structure of the interaction at the
vertex is illustrated in Fig. 1, in which horizontal lines correspond to
real gluons produced in the intermediate states and vertical and inclined  lines
describe  propagating reggeized gluons.
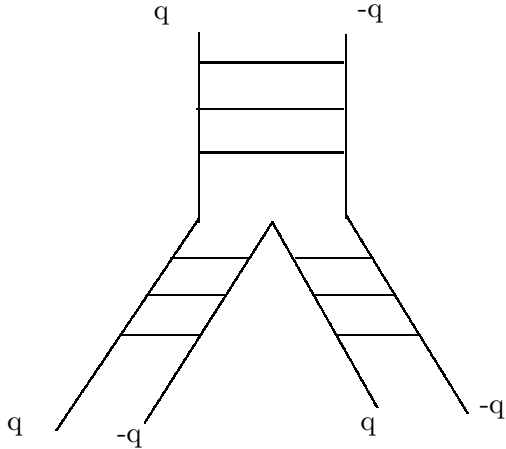
\begin{figure} [ht]
\unitlength 1mm 
\linethickness{0.4pt}
\ifx\plotpoint\undefined\newsavebox{\plotpoint}\fi 
\begin{picture}(100.25,103.75)(0,0)
\put(61.25,101.25){\line(0,-1){25.25}}
\put(80.75,101){\line(0,-1){24.5}}
\multiput(61.25,76.5)(-.033687943,-.050088652){564}{\line(0,-1){.050088652}}
\put(42.25,48.25){\line(0,1){0}}
\put(80.75,76.75){\line(1,0){.25}}
\multiput(81,76.75)(-.03125,.03125){8}{\line(0,1){.03125}}
\multiput(80.75,77)(.033673469,-.055102041){245}{\line(0,-1){.055102041}}
\multiput(89,63.5)(.03361345,-.05462185){238}{\line(0,-1){.05462185}}
\put(71,75.75){\line(0,1){.25}}
\multiput(71,76)(-.033730159,-.053075397){504}{\line(0,-1){.053075397}}
\multiput(71,76)(.03373494,-.059638554){415}{\line(0,-1){.059638554}}
\put(61.25,97.25){\line(1,0){19.25}}
\put(61,91){\line(1,0){19.5}}
\put(61.25,85.25){\line(1,0){19.25}}
\put(57.5,71.25){\line(1,0){10.25}}
\put(59.25,65){\line(0,1){0}}
\put(54.25,66.25){\line(1,0){10.5}}
\put(51,61){\line(1,0){10.25}}
\put(74,71.25){\line(1,0){10}}
\put(76.5,66.25){\line(1,0){10.5}}
\put(79.5,61){\line(1,0){11}}
\put(56.25,103.5){\makebox(0,0)[cc]{q}}
\put(36.75,48.75){\makebox(0,0)[cc]{q}}
\put(84,103.75){\makebox(0,0)[cc]{-q}}
\put(100.25,51){\makebox(0,0)[cc]{-q}}
\put(52,47.25){\makebox(0,0)[cc]{-q}}
\put(83.75,48.75){\makebox(0,0)[cc]{q}}
\end{picture}
\vspace*{-4 cm}

\caption{Interaction of three BFKL pomerons at the
spliiting vertex} \label{Fig1}
\end{figure}
From this structure one sees that
the produced gluons are contained in the intermediate states of the interacting
pomerons, so that to get the inclusive cross-section one has to "open" these
pomerons, that is to fix the momentum of one of the intermediate real gluons
in them. For pA collisions this leaves the diagrams of the type shown
in Fig 2$a$.
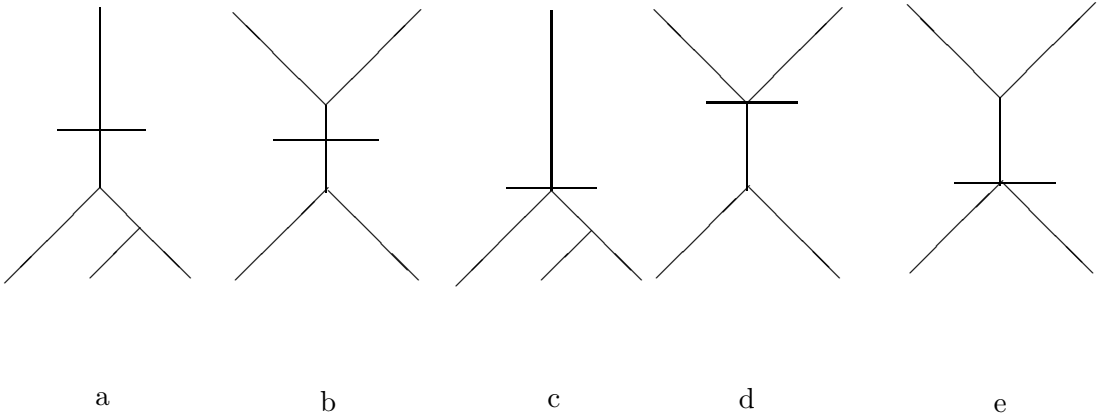
\begin{figure}
\unitlength=1.00mm
\special{em:linewidth 0.4pt}
\linethickness{0.4pt}
\begin{picture}(148.67,110.67)
\put(16.33,110.00){\line(0,-1){24.00}}
\put(16.33,86.00){\line(-1,-1){12.67}}
\put(16.33,86.00){\line(1,-1){12.00}}
\put(21.67,80.67){\line(-1,-1){6.67}}
\put(34.33,73.67){\line(1,1){12.33}}
\put(46.67,85.67){\line(1,-1){12.00}}
\put(46.33,85.33){\line(0,1){11.67}}
\put(46.33,97.00){\line(-1,1){12.33}}
\put(46.33,97.00){\line(1,1){12.67}}
\put(90.33,74.00){\line(1,1){12.33}}
\put(102.67,86.00){\line(1,-1){12.00}}
\put(102.33,85.67){\line(0,1){11.67}}
\put(102.33,97.33){\line(-1,1){12.33}}
\put(102.33,97.33){\line(1,1){12.67}}
\put(124.00,74.67){\line(1,1){12.33}}
\put(136.33,86.67){\line(1,-1){12.00}}
\put(136.00,86.33){\line(0,1){11.67}}
\put(136.00,98.00){\line(-1,1){12.33}}
\put(136.00,98.00){\line(1,1){12.67}}
\put(10.67,93.67){\line(1,0){11.67}}
\put(39.33,92.33){\line(1,0){14.00}}
\put(97.00,97.33){\line(1,0){12.00}}
\put(130.00,86.67){\line(1,0){13.33}}
\put(16.66,58.00){\makebox(0,0)[cc]{a}}
\put(46.66,57.67){\makebox(0,0)[cc]{b}}
\put(76.33,109.67){\line(0,-1){24.00}}
\put(76.33,85.67){\line(-1,-1){12.67}}
\put(76.33,85.67){\line(1,-1){12.00}}
\put(81.67,80.34){\line(-1,-1){6.67}}
\put(70.33,86.00){\line(1,0){12.00}}
\put(76.67,57.67){\makebox(0,0)[cc]{c}}
\put(102.33,58.00){\makebox(0,0)[cc]{d}}
\put(136.00,57.00){\makebox(0,0)[cc]{e}}
\end{picture}

\vspace*{-5cm}
\caption{Typical diagrams for the inclusive cross-section in nucleus-nucleus
collisions.}
\label{Fig2}
\end{figure}
For the nucleus-nucleus case
a similar production mechanism in the old-fashioned local pomeron
model was proven to lead to inclusive cross-sections given by
a convolution of two sets of fan diagrams connecting
the emitted particle to the two nuclei times the vertex for
the emission (Fig. 2$b$). The proof was based on the
AGK rules  appropriately adjusted for the
triple pomeron interaction ~\cite{cia}.
It was later shown  in ~\cite{BW} that the AGK rules
are fulfilled for interacting BFKL pomerons.
So the same arguments as in ~\cite{cia}
allow to demonstrate that for collisions of two nuclei
the inclusive cross-section will
be given by  the same Fig.2$b$, that is, apart from the emission
vertex,  by the convolution of two sums of fan diagrams,
constructed of BFKL pomerons and triple pomeron verteces, propagating
from the emitted particle towards the two nuclei ~\cite{bra4}.

Taking into account the form of the emission vertex (see ~\cite{bra4})
 we obtain for the pA case
\beq
I_{A}(y,k)=\frac{8N_c\alpha_s}{k^2}\int d^2\beta
d^2re^{ikr}\Delta\Psi(Y-y,r)\cdot
\Delta\Phi(y,r,\beta),
\eeq
and for the AA case at $b=0$ \footnote{A sligtly different coefficient
as compared to ~\cite{bra5} is due to different normalization of $\Phi$.}
\beq
I_{AA}(y,k)=\frac{N_c\alpha_s}{\pi^2\alpha_{s0}^2k^2}\int d^2\beta
d^2re^{ikr}\Delta  \Phi(Y-y,r,\beta)\cdot
\Delta  \Phi(y,r,\beta).
\eeq
Here  $\Psi(y,r)$ is a pomeron at rapidity $y$ and of the
transverse dimension $r$ coupled to the incoming
proton.  Its normalization at $y=0$ is
\beq
\Psi(0,r)=\frac{2\Phi(0,r,b)}{g_0^4AT_A(b)}
\eeq
(obviously the right-hand side of (16) does not depend on $b$).
$\Delta$'s are two-dimensional Laplacians applied to $\Psi$
and $\Phi$'s.

Later from the colour dipole formalism a slightly different form for
the inclusive cross-section was derived in ~\cite{KT}. For the
dipole-nucleus scattering case it corresponds to changing
\beq
2\Phi(y,\beta,r)\to 2\Phi(y,\beta,r)-\Phi^2(y,\beta,r).
\eeq
Note that in ~\cite{KT} it was erroneously stated that the change
was from the "quark dipole"  $\Phi$ to the "gluon dipole"
$2\Phi-\Phi^2$. As seen from ( 17 ) it is not. In fact the change is
equivalent to adding to the AGK contribution ( 15 ) a new one which
has the meaning of the emission of the gluon from the triple pomeron
vertex itself. Such a contribution is not prohibited in principle.
From our point of view, taking into account the structure of the
vertex shown in Fig. 1, its appearance is difficult to understand.
However in this paper we do not pretend to discuss the validity of
the two proposed formulas for the inclusive cross-sections on the
fundamental level. Rather we shall compare the cross-sections which
follow from them after numerical calculations.

For the pA case this recipe implies taking into account a new diagram
shown in Fig. 2 $c$.
As a result, one finds, instead of ( 14 ), the Kovchegov-Tuchin (KT)
cross-section
\beq
I^{KT}_{A}(y,k)=\frac{4N_c\alpha_s}{k^2}\int d^2\beta
d^2re^{ikr}\Delta\Psi(Y-y,r)\cdot
\Delta[2\Phi(y,r,\beta)-\Phi^2(y,r,\beta)],
\eeq

For the nucleus-nucleus case the recipe of ~\cite{KT} implies taking into
account two
new diagrams for the inclusive cross-sections shown in Fig. 2 $d$ and $e$.
The nucleus-nucleus cross-section thus becomes

\[
I_{AA}^{KT}(y,k)=\frac{N_c\alpha_s}{2\pi^2\alpha_{s0}^2k^2}\int d^2\beta
d^2re^{ikr}\Big[2\Delta\Phi(Y-y,r,\beta)\Delta\Phi(y,r,\beta)-\]\beq
\Delta\Phi(Y-y,r,\beta)\Delta\Phi^2(y,r,\beta)-
\Delta\Phi^2(Y-y,r,\beta)\Delta\Phi(y,r,\beta)\Big].
\eeq

Passing in our formulas to the momentum space we find
(suppressing the dependence on $y$ and $\beta$)
\beq
\Delta\Phi(r)\to 2\pi q^2\Delta_q\phi(q)\equiv 2\pi h(q)
\eeq
Introducing also a function similar to $h$ for the pomeron $\Psi$:
\beq
\Delta\Psi(r)=2\pi h^{(0)}(q)
\eeq
we can express the cross-sections (14) and (15) via the gluon distributions
$h^{(0)}(q)$ and $h(q)$ in a factorized form:
\beq
I_{A}(y,k)=\frac{8N_c\alpha_s}{k^2}\int d^2\beta
d^2q h^{(0)}(Y-y,q)
h(y,k-q,\beta),
\eeq
and for the AA case at $b=0$
\beq
I_{AA}(y,k)=\frac{N_c\alpha_s}{\pi^2\alpha_{s0}^2k^2}\int d^2\beta
d^2qh(Y-y,q,\beta)
h(y,k-q,\beta).
\eeq

The KT cross-section contains an additional term
\beq
X(r)=-\Delta \Phi^2(r)=-\Delta\, (2\pi)^2r^4\phi^2(r).
\eeq
We transform it to the momentum space:
\beq
X(q)=q^2\int d^2re^{-iqr}\Phi^2(r)=
(2\pi)^2q^2\int d^2r \Big(r^2\phi(r)\Big)^2.
\eeq
On the other hand
\beq
r^2\phi(r)=-\int \frac{d^2q'}{(2\pi)^2}e^{iq'r}\Delta'\phi(q').
\eeq
So we get
\beq
X(q)=
q^2\int d^2q_1\Delta
\phi(q_1)\cdot\Delta\phi(q-q_1).
\eeq
Taking into account that
\beq
\Delta\phi(q)=-2\pi\delta(q)+\frac{h(q)}{q^2}
\eeq
gives
\beq
X=-4\pi h(q)+q^2\int\frac{d^2q_1}{q_1^2(q-q_1)^2}h(q_1)h(q-q_1).
\eeq

This should be added to the first term in the 2nd factor of the
integrand in (18),
which according to (20) is just
$4\pi h(q)$. So in the end  in the KT cross-section
instead of $4\pi h(q)$ will appear
the second term in ( 29 ). Introducing the function
\beq
w(q)=\frac{q^2}{2\pi}\int\frac{d^2q_1}{q_1^2(q-q_1)^2}h(q_1)h(q-q_1)
\eeq
we finally find that the KT cross-section is obtained from the AGK cross-section
by the substitution
$
h(q)\to w(q)/2
$:
\beq
I^{KT}_{A}(y,k)=\frac{4N_c\alpha_s}{k^2}\int d^2\beta
d^2q h^{(0)}(Y-y,q)
w(y,k-q,\beta),
\eeq

Now we pass to the nucleus-nucleus cross-section.
For different colliding nuclei the KT rule means (see Fig. 2)
\beq
2\Delta\Phi_A(r)\cdot \Delta\Phi_B(r)\to
2\Delta\Phi_A(r)\cdot \Delta\Phi_B(r)-\Delta\Phi_A(r)\cdot\Delta\Phi_B^2(r)-
\Delta\Phi_A^2(r)\cdot\Delta\Phi_B(r)
\eeq
where we again suppress all
other arguments in $\Phi$ evident from (19).
Going to  the momentum space, we have
\beq
\Delta\Phi(r)\to 2\pi h(q),\ \ -\Delta\Phi^2(r)\to -4\pi h(q)+2\pi w(q)
\eeq
and we shall find the KT cross-section at a given impact parameter $\beta$
as
\[
I_{AB}^{KT}(y,k,\beta)=
\frac{N_c\alpha_s}{2\pi^2\alpha_{s0}^2k^2}
\int d^2bd^2q\Big(h_A(Y-y,q,\beta-b)w_B(y,k-q,b)\]\beq+
w_A(Y-y,q,\beta-b)h_B(y,k-q,b)-2h_A(Y-y,q,\beta-b)h_B(y,k-q,b)\Big).
\eeq
For central collisions of identical nuclei at mid-rapidity this simplifies to
\beq
I_{AA}^{KT}(Y/2,k,0)=
\frac{N_c\alpha_s}{\pi^2\alpha_{s0}^2k^2}
\int d^2bd^2qh(Y/2,q,b)\Big(w(Y/2,k-q,b)-
h(Y/2,k-q,b)\Big).
\eeq
So in the end the KT cross-sections can also be expressed via function
$h(q)$.

Function $h(y,k,\beta)$ has a normalization property ~\cite{bra4}
\beq
 \int \frac{d^2k}{k^2}h(y,k,\beta)=1 \eeq and at sufficiently
high $y$  acquires a scaling property \beq
h(y,k\beta)=h\Big(k/Q_s(y,\beta)\Big),
\eeq
where $Q_s(y,\beta)$
is the above-mentioned saturation momentum. From ( 36 ) and ( 37 )
one easily establishes some properties of the new function $w_A$.
Obviously it scales with the same saturation momentum when $h_A$
does
\beq
 w(y,k\beta)=w\Big(k/Q_s(y,\beta)\Big).
\eeq
 At
$k\to\infty$ it has the asymptotic
 \beq
w(y,k,\beta)_{k\to\infty}\sim 2h(y,k,\beta)
\eeq
 and finally
\beq
\int d^2k w(y,k,\beta)=2\int d^2k h(y,k,\beta).
\eeq

These properties immediately allow to make some preliminary comparison
between the cross-sections given by ( 22 ) and ( 23 ), on the one hand, and
( 31 ) and ( 34 ) , on the other. Obviously if $k/Q_s$ is large
both expressions give the same cross-section due to ( 40 ). In the opposite
limit of small $k/Q_s$, the scaling property allows to conclude
that the ratio of the two
cross-sections is a universal constant which does not depend on $y$,
nor on $A$
nor on $\beta$. Our numerical results  confirm these predictions.

\section{Results}
Our cross-sections  depend on three parameters, $\sigma_0$  and two
coupling constants $\alpha_s$ at the comparatively large
evolution scale and $\alpha_{s0}$
at the very low proton scale.
As mentioned, we have certain control on the
value of $\sigma_0$ from the experimental data on the proton structure
function at low $x$, embodied in the parametrization (3) borrowed
from ~\cite{gobi}. However we can have only very general ideas about
possible values
for the coupling constant at the two scales involved, since the coupling constant
is  not running in the model. The common wisdom is
to take $\alpha_s=0.2$, which is considered to be the value reached
by the running coupling at relatively low scales. In our calculations
we have taken this value for the evolution equation and for the
production vertex.
However with the same value for $\alpha_{s0}$ we
find the total pp and pA cross-sections
at low energies roughly 4 times greater than the
experimental data.
To bring them into the physically reasonable
range we have to double the value of $\alpha_{s0}$.
So in our claculations we have taken $\alpha_{s0}=0.4$.

\subsection{The gluon densities and saturation momentum}
The gluon density in the nucleus is , up to a numerical coefficient,
given by the function $h(y,k,b)$ 
introduced in (7) ~\cite{bra1}:
\beq
\frac{d[xG(x,k^2,b)]}{d^bd^2k}=\frac{2N_c}{\pi g^2}h(y,k,b),\ \
y=\ln \frac{1}{x}.
\eeq
The role of $h$ as the gluon density is clearly illustrated in the
factorized forms of the inclusive cross-sections ( 22 ) and ( 23 ) for
pA and AA collisions. In the KT  inclusive cross-section for hA collisions
function $h(y,k,b)$ is however substituted by  $(1/2)w(y,k,b)$, which can
also be considered an effective gluon density taking into account
the contribution from the emission vertex. Both densities are well defined
for fixed $b$. As mentioned, at $\bar{y}\geq 1$ they acquire a
scaling behaviour and depend only on the ratio $k/Q_s(y,b)$ where
$Q_s$ is the saturation momentum. In the minimum bias pA collisions and
in AB collisions at
a fixed impact parameter they are smeared out over the nucleus transverse
space. In Fig. 3 we illustrate the gluon densities $h$ and $w$ at
$b=0$ for A=180 and $\bar{y}=2$, which, with our value for $\alpha_s$,
corresponds to $y\sim 10$.
\begin{figure}[ht]
\epsfxsize 4in
\centerline{\epsfbox{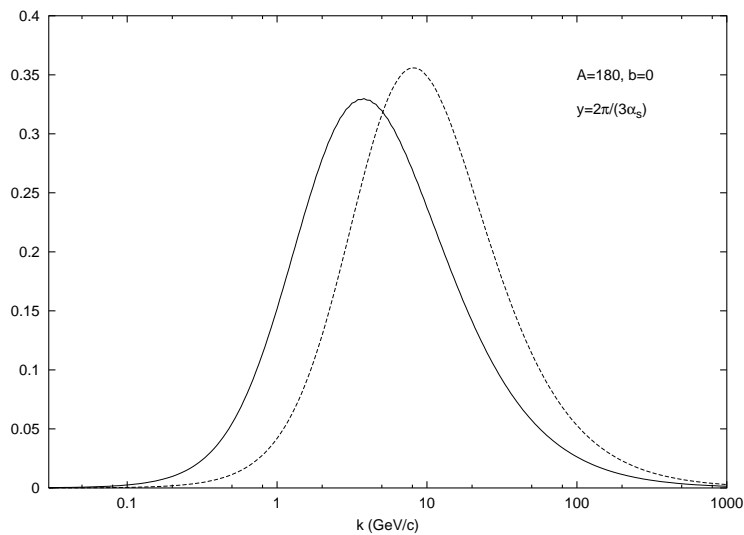}}
\caption{Gluon densities $h(k)$ (the left curve) and $w(k)$
for A=180 at $b=0$ and $\bar{y}=2$}
\label{Fig3}
\end{figure}
As one observes, both functions have the same form, with a clear maximum
at a certain value of $k$. This value for $h$ is the value
for the saturation momentum $Q_s$ ~\cite{arbra}.
For $w$ the maximum is somewhat shifted towards higher momenta
and the value of $(1/2)w$ at the maximum is lower than that for $h$.
As is clear from our formulas for the inclusive
cross-sections, these two properties  have opposite effect, so that  in
the end
the inclusive cross-sections derived from AGK and introduced by KT result
practically equal.  In Fig. 4 we show the dependence of the  saturation
momentum $Q_s$ on $b$ for $A=180$ and at $\bar{y}=2$.
\begin{figure}[ht]
\epsfxsize 4in
\centerline{\epsfbox{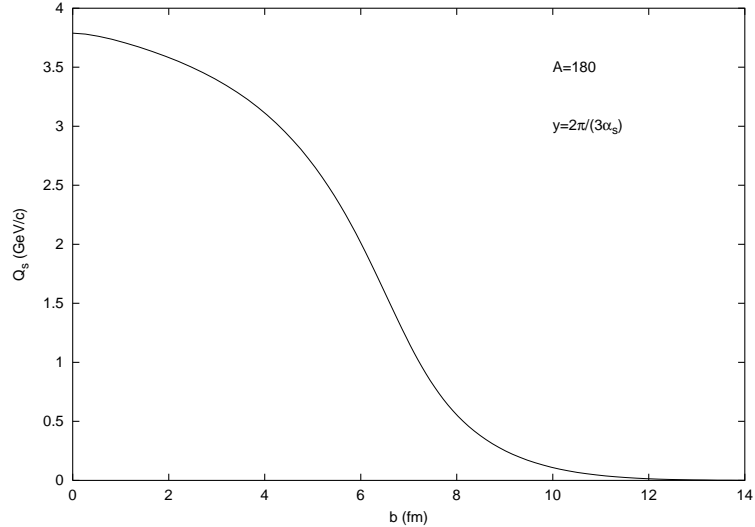}}
\caption{The saturation momentum $Q_s$ as a function of $b$
for A=180 at $\bar{y}=2$}
\label{Fig4}
\end{figure}
Its form closely follows the nuclear density $T_A(b)$ reaching
3.8 GeV/c at the center and dropping to 0.25 GeV/c  and lower at
its periphery.

\subsection{pA cross-sections}
Passing to pA cross-sections we start with the total cross-sections
(Eq. (8)). They are shown in Fig.5 as a function of $y$ for $A=9,27,
64,108$ and 180, divided by $A^{2/3}$ to compare with  purely geometric
cross-sections.
\begin{figure}[ht]
\epsfxsize 4in
\centerline{\epsfbox{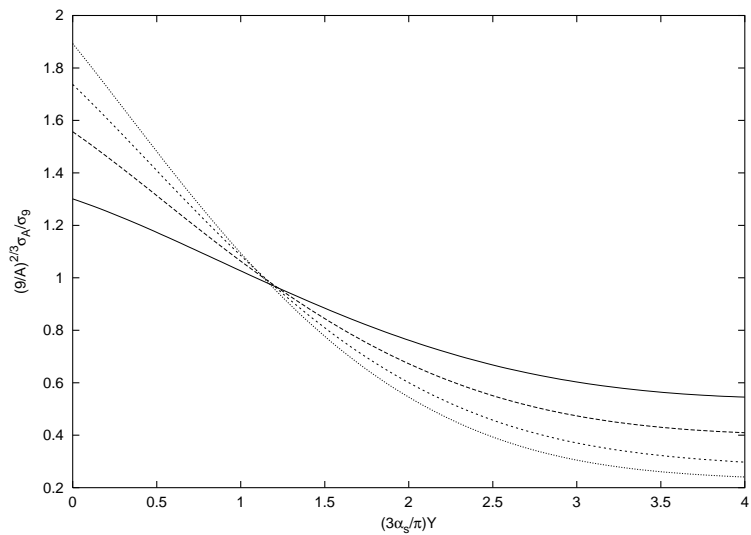}}
\caption{The total pA cross-sections divided by $A^{2/3}$. Curves
from top to bottom on the left correspond to $A=9,27,64,108$ and 180}
\label{Fig5}
\end{figure}
As one observes, the cross-sections become strictly geometric at
$\bar{y}\simeq 1.2$. At lower energies they grow faster than $A^{2/3}$,
at higher energies they grow much slowlier, approximately as $A^{0.2}$.
This can be explained by the role of peripheric parts of the nucleus,
whose contribution grows with energy due to absence of non-linear damping
effects and is relatively greater in lighter nuclei.

Inclusive cross-sections were calculated at the overall rapidity
$\bar{Y}=4$, which correspond to the natural rapidity $\sim 20$.
Inclusive cross-sections corresponding to Eq. (14) are shown in Figs. 6-8.
Absolute values for the inclusive
cross-sections are presented in Fig.6 for $A=108$ at central rapidity
($y=Y/2$). At $k< Q_s$ they are $\propto 1/k^2$, so that the integral
factor in Eq. (14) is practically constant. At $k>Q_s$ the cross-sections fall
more rapidly. At $k\gg Q_s$ they behave as $\sim 1/k^{3.1}$. However this
asymptotic is only reached at very high nomenta $\sim 10^5$ GeV/c.
In Fig. 7 we show the inclusive cross-sections for $A=108$
at different rapidities relative to the central rapidity. As expected,
the cross-section grow towards the projectile  region, since the rapidity of
the pure pomeron factor is growing in this region.
\begin{figure}[ht]
\epsfxsize 4in
\centerline{\epsfbox{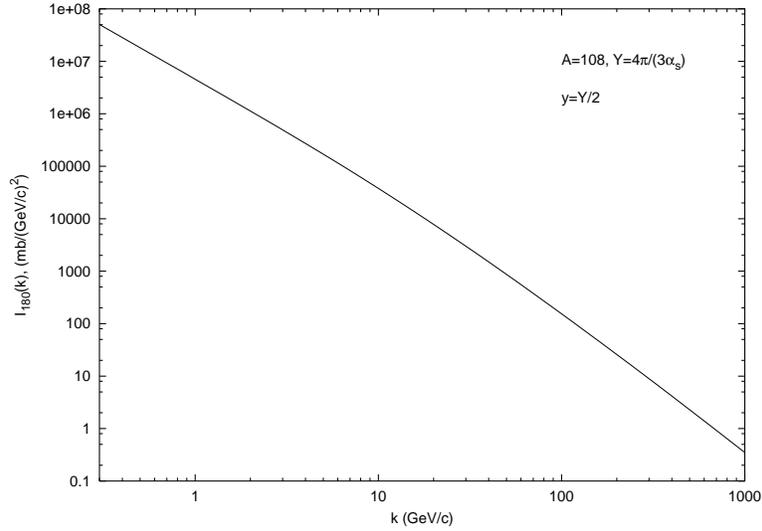}}
\caption{Inclusive pA cross-sections for $A=108$ at $\bar{Y}=4$
and $y=Y/2$}
\label{Fig6}
\end{figure}
\begin{figure}[ht]
\epsfxsize 4in
\centerline{\epsfbox{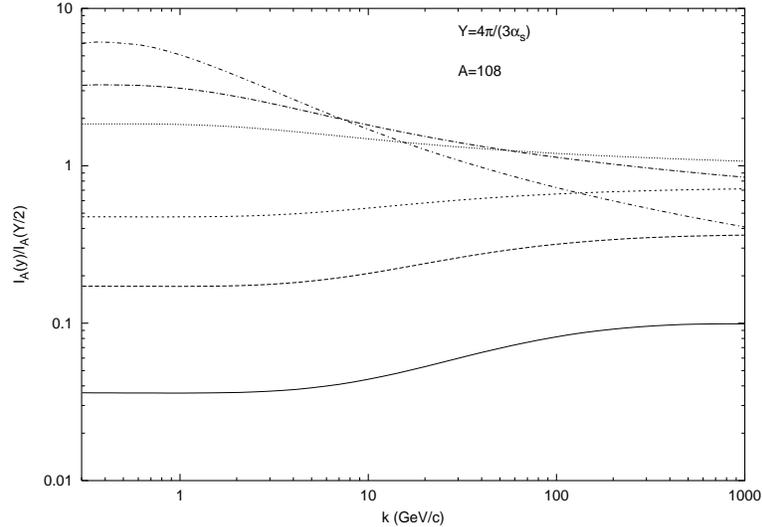}}
\caption{Inclusive pA cross-sections for $A=108$ at $\bar{Y}=4$
and different $y$ relative to $y=Y/2$.
Curves from top to bottom correspond to $y/Y=1/8,1/4,3/8,5/8,3/4$ and 7/8}
\label{Fig7}
\end{figure}
The $A$-dependence of the inclusive cross-sections is different at
low momenta, below $Q_s$ and high momenta, much larger than $Q_s$.
This is illustrated in Fig.8 where we plot
$ (9/A)^{0.745}I_A(k)/I_9(k)$.
\begin{figure}[ht]
\epsfxsize 4in
\centerline{\epsfbox{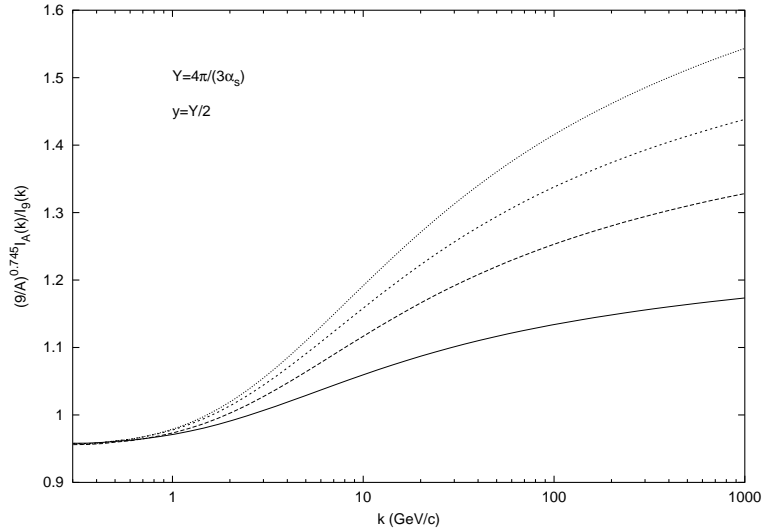}}
\caption{The $A$-dependence of  inclusive pA cross-sections
at $\bar{Y}=4$ and $y=Y/2$. Curves from bottom to top on the right
correspond to $A=27,64,108$ and 180}
\label{Fig8}
\end{figure}
It is clearly seen that at low momenta the cross-sections are
$\propto A^{0.745}$ whereas at high momenta they grow faster with $A$,
approximately as $A^{0.9}$. In any case their growth with $A$ is damped as
compared with naive probabilistic expectations  which predict them
to be $\propto A$.

Passing to the determination of multiplicities one has to observe
certain care because of the properties of the perturbative QCD
solution in the leading approximation embodied in Eqs. ( 14 ) and ( 15 ).
As follows from these formulas  inclusive cross-sections blow up at
$k^2\to 0$ independently of rapidity $y$. So the corresponding
total multiplicity diverges logarithmically. However, the physical
sense has only emission of jets with high enough transverse momenta.
Thus one has to cut the spectrum from below by some $k_{min}$ which
separates the spectrum of jets proper from soft gluons which are not
related to jets. Inevitably the multiplicity of thus defined jets
depends on the  chosen value of $k_{min}$. We have chosen
$k_{min}=2$ GeV/c.  In Fig.9 we show multiplicities at $\bar{Y}=2$
divided by $A^{2/3}$ as a function of the rapidity $y$. One observes
that on the whole the multiplicities are roughly $\propto A^{2/3}$.
They diminish with $y$, except at very small $y$, which is natural,
since the non-linear damping of the gluon density grows as one goes up
along the fans.
\begin{figure}[ht]
\epsfxsize 4in
\centerline{\epsfbox{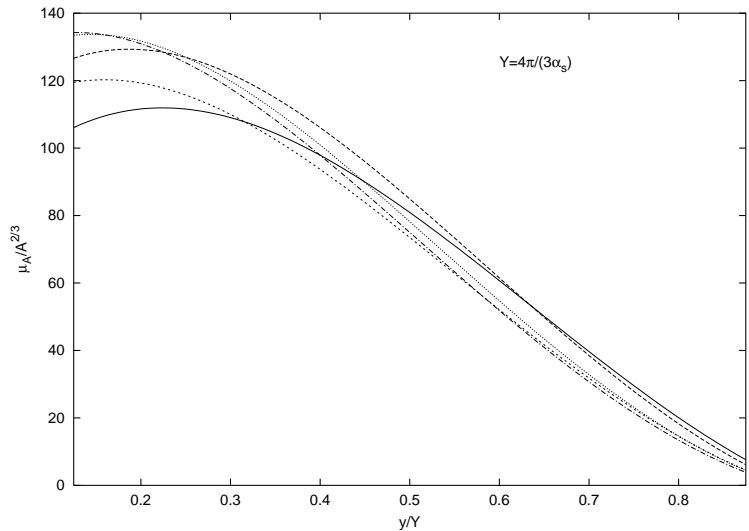}}
\caption{pA multiplicities  at $\bar{Y}=4$, divided by $A^{2/3}$.
Curves from  top to bottom on the right correspond to
$A=9,27,64,108$ and 180}
\label{Fig9}
\end{figure}

Finally we pass to the cross-sections obtained with the KT formula
( 18 ). In Fig. 10 we show  ratios of these cross-sections to the
ones defined by the AGK rules, Eq. ( 14 ), for $\bar{Y}=4$ and
$y=Y/4,Y/2$ and $(3/4)Y$. These ratios turn to unity at $k$
above $Q_s$, as discussed in the end of the preceding
section. Below $Q_s$ the ratios are in the region $0.72\div
0.87$ with little dependence on $A$. Some dependence which is
left can be explained by the fact that for the BFKL pomeron
attached to the projectile the scaling regime is not valid and
that for very peripheral parts of
the nucleus it can only be reached at rapidities
well above the considered ones. Due to this very simple relation
between the two cross-sections, all conclusions about the $A$
-dependence drawn for the AGK cross-section ( 14 ) remain valid also
for the KT cross-section ( 18 ).
\begin{figure}[ht]
\epsfxsize 4in
\centerline{\epsfbox{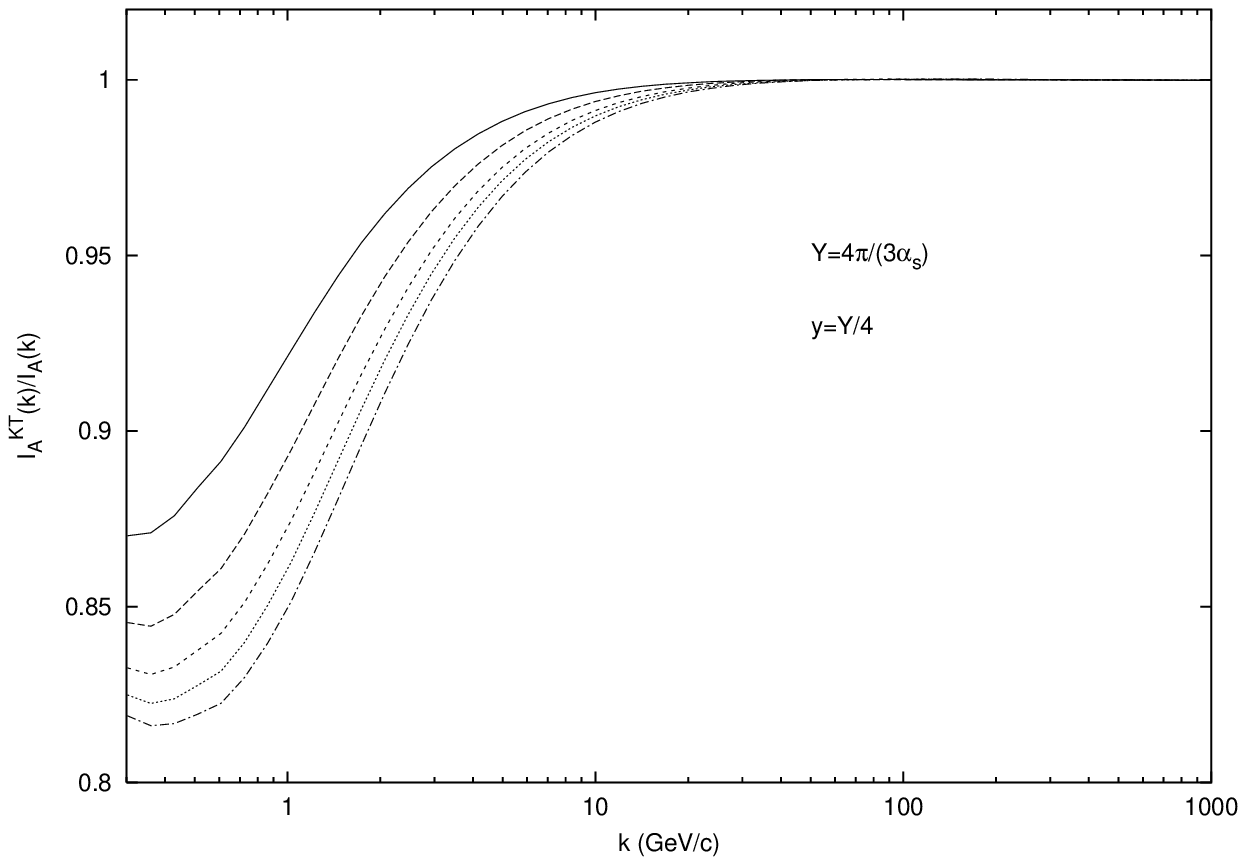}}
\epsfxsize 4in
\centerline{\epsfbox{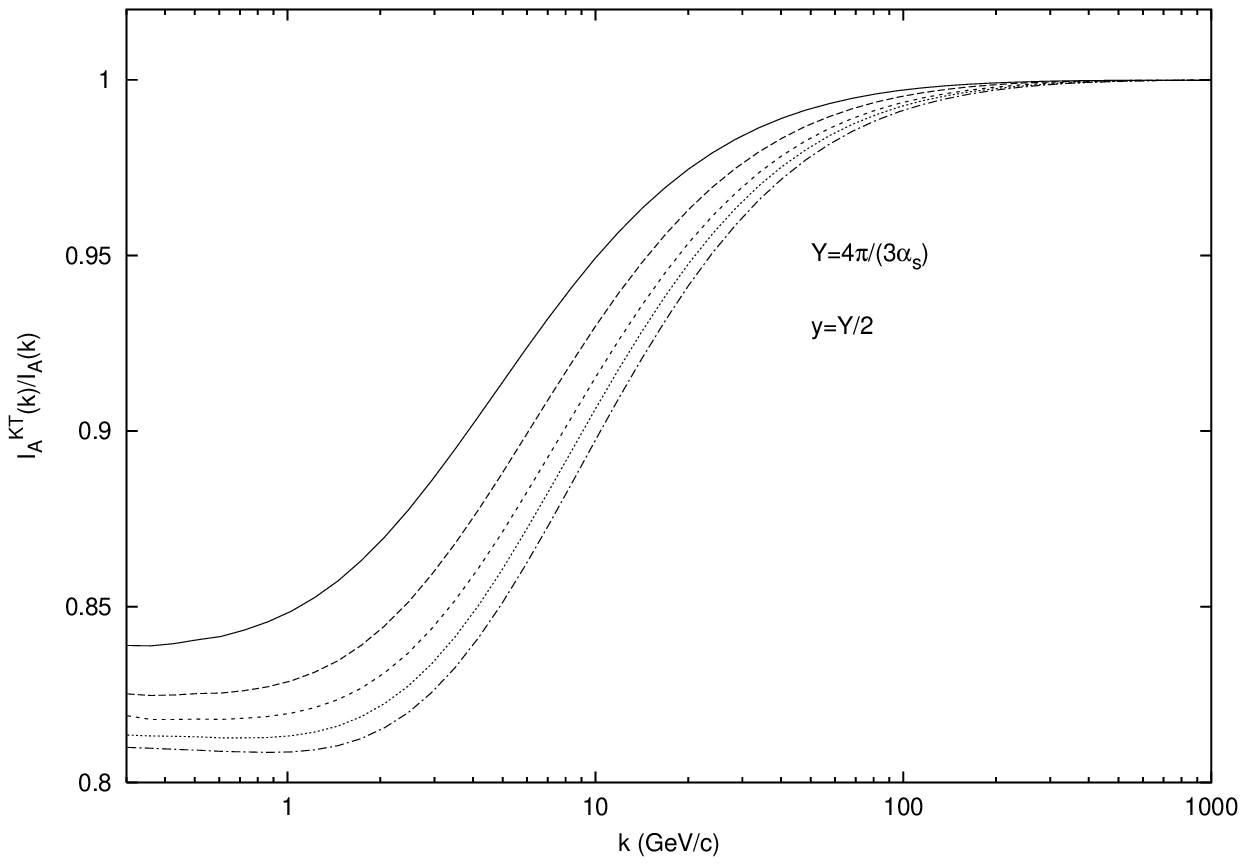}}
\epsfxsize 4in
\centerline{\epsfbox{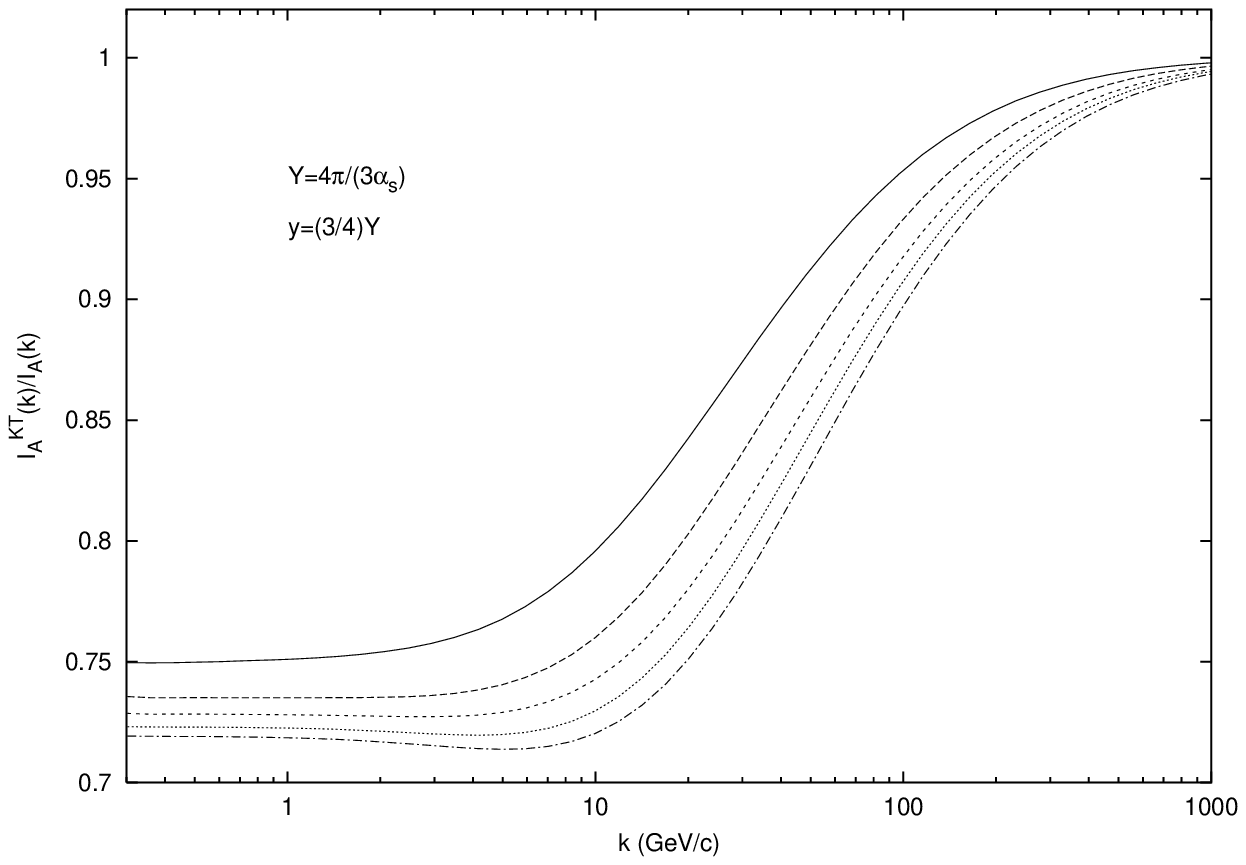}}
\caption{Ratios of the KT inclusive pA cross-sections, Eq. ( 18 ), to the
ones found from the AGK rules, Eq. ( 14 ) at $\bar{Y}=4$.
Curves from top to bottom refer to $A=9, 27, 64, 108$ and 180 }
\label{Fig10}
\end{figure}

\subsection {AA collisons}
As mentioned, we only considered central collisions of identical nuclei
($A=B$ and $b=0$). The inclusive cross-sections obtained
from the AGK rules (Eq. (15)) are shown in Figs. 11-13. Fig. 11 presents
absolute values for the inclusive cross-sections
for $A=108$ at center rapidity and different overall energies
corresponding to $\bar{Y}=2,4$ and 8.  As for pA collisions, at $k<Q_s$
their behaviour is totally determined by the $1/k^2$ factor in Eq. ( 19 ),
the integral factor being practically independent of $k$. At very
high momenta
$I_A(Y,y=Y/2,k)\sim 1/k^{p(y)}$ with power $p(y)$ diminishing with energy.
From our calculations we find that $p(y)\simeq 3.3,\,3.0$ and 2.7
at $\bar{y}=2,\,4$ and 8 respectively. At infinite energies $p$ seems to tend
to 2 in correspondence with $Q_s\to\infty$.
\begin{figure}[ht]
\epsfxsize 4in
\centerline{\epsfbox{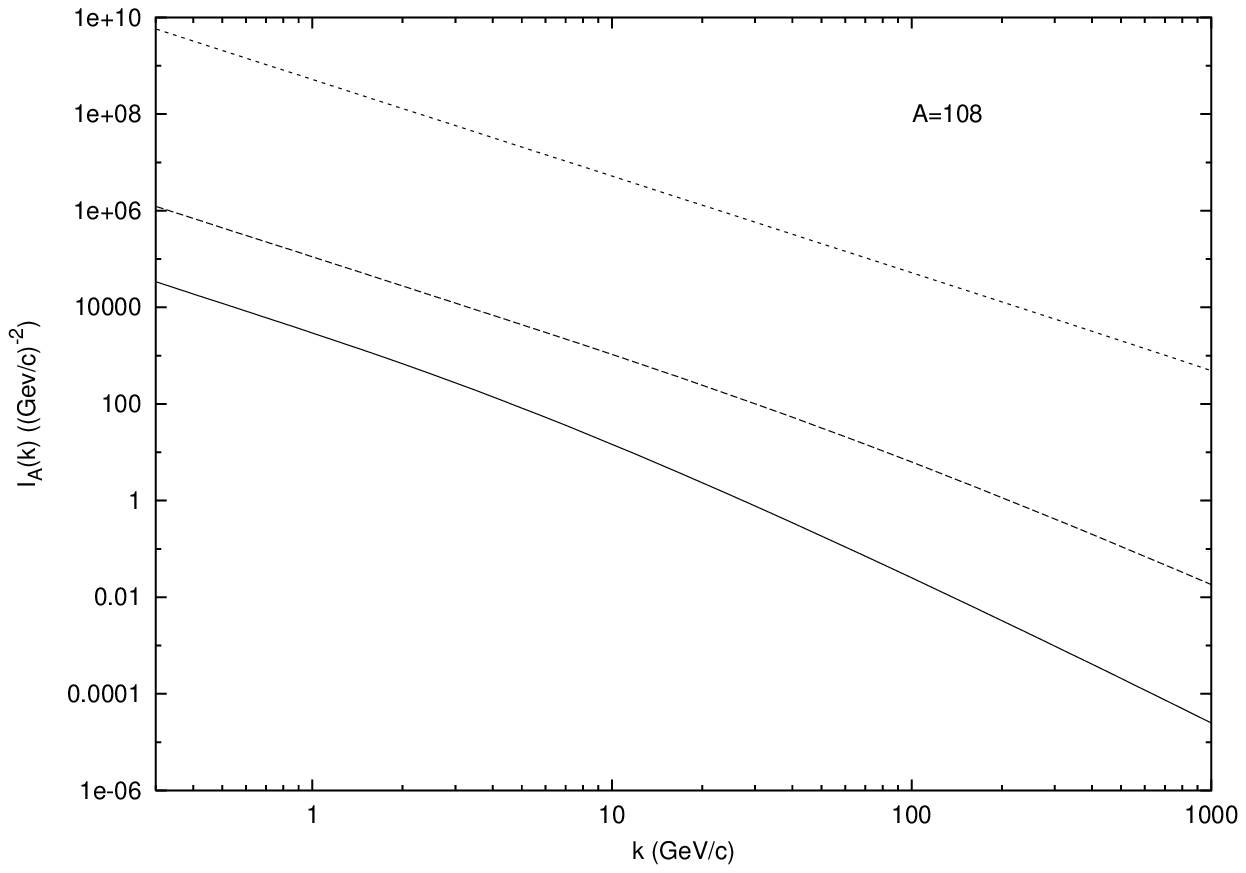}}
\caption{AA inclusive cross-section at mid-rapidity (y=Y/2) for
$A=180$
Curves from  top to bottom correspond to the overall rapidities
$\bar{Y}=8,4$ and 2}
\label{Fig11}
\end{figure}
To see the dependence on the rapidity $y$ of
produced particles, in Fig.12 we show the  cross-sections at different
$y$ for $A=108$ and the overall rapidity $\bar{Y}=4$. As expected
the cross-sections steadily diminish towards the edges, their form not
changing seriously.
\begin{figure}[ht]
\epsfxsize 4in
\centerline{\epsfbox{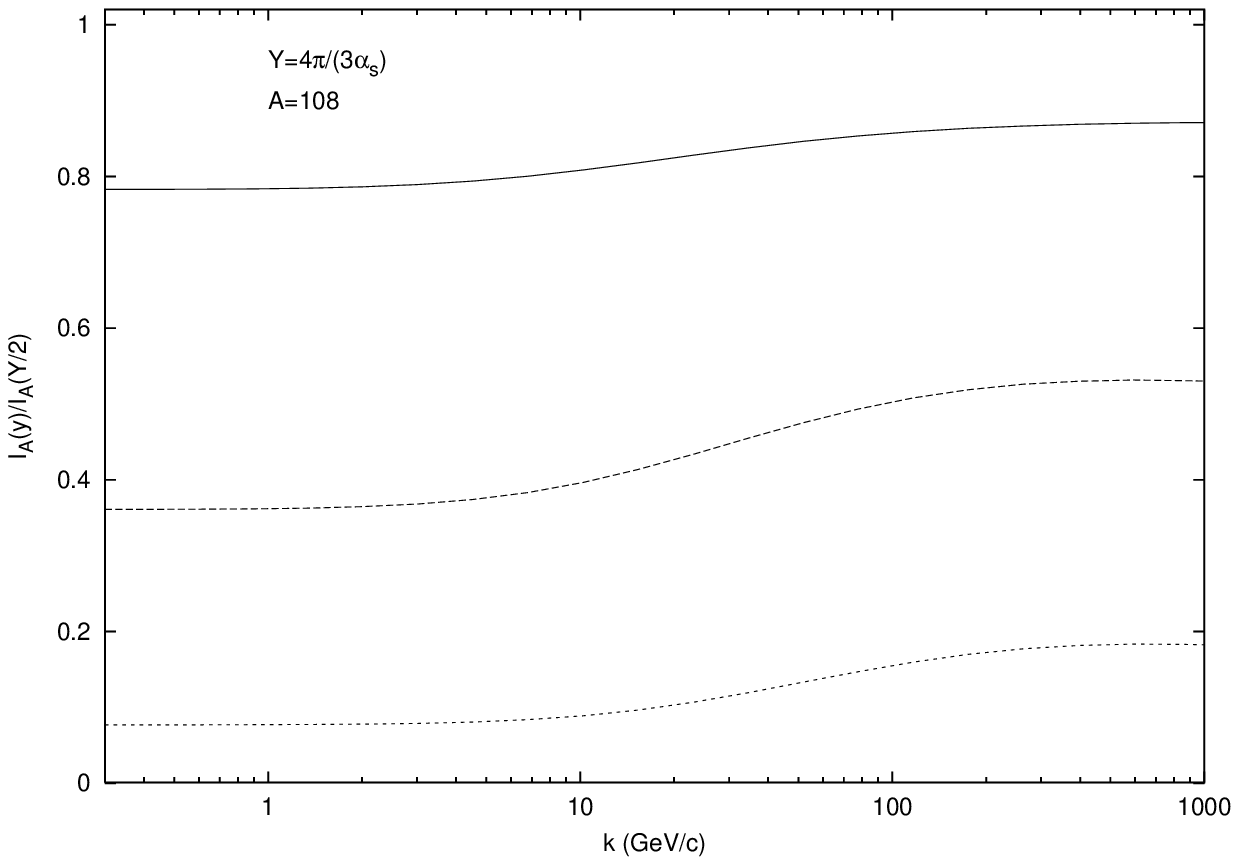}}
\caption{AA inclusive cross-setions at different $y$ for
$A=180$ and $\bar{Y}=4$ relative to  midrapidity $y=Y/2$.
Curves from  top to bottom correspond to
$y=(3/8)Y,(1/2)Y$ and $(1/8)Y$}
\label{Fig12}
\end{figure}
The $A$-dependence of the cross-sections is illustrated in Fig. 13
where we show the ratio
$ (9/A)I_{A}/I_9(k)$
at mid-rapidity for the overall rapidity $\bar{Y}=4$.
\begin{figure}[ht]
\epsfxsize 4in
\centerline{\epsfbox{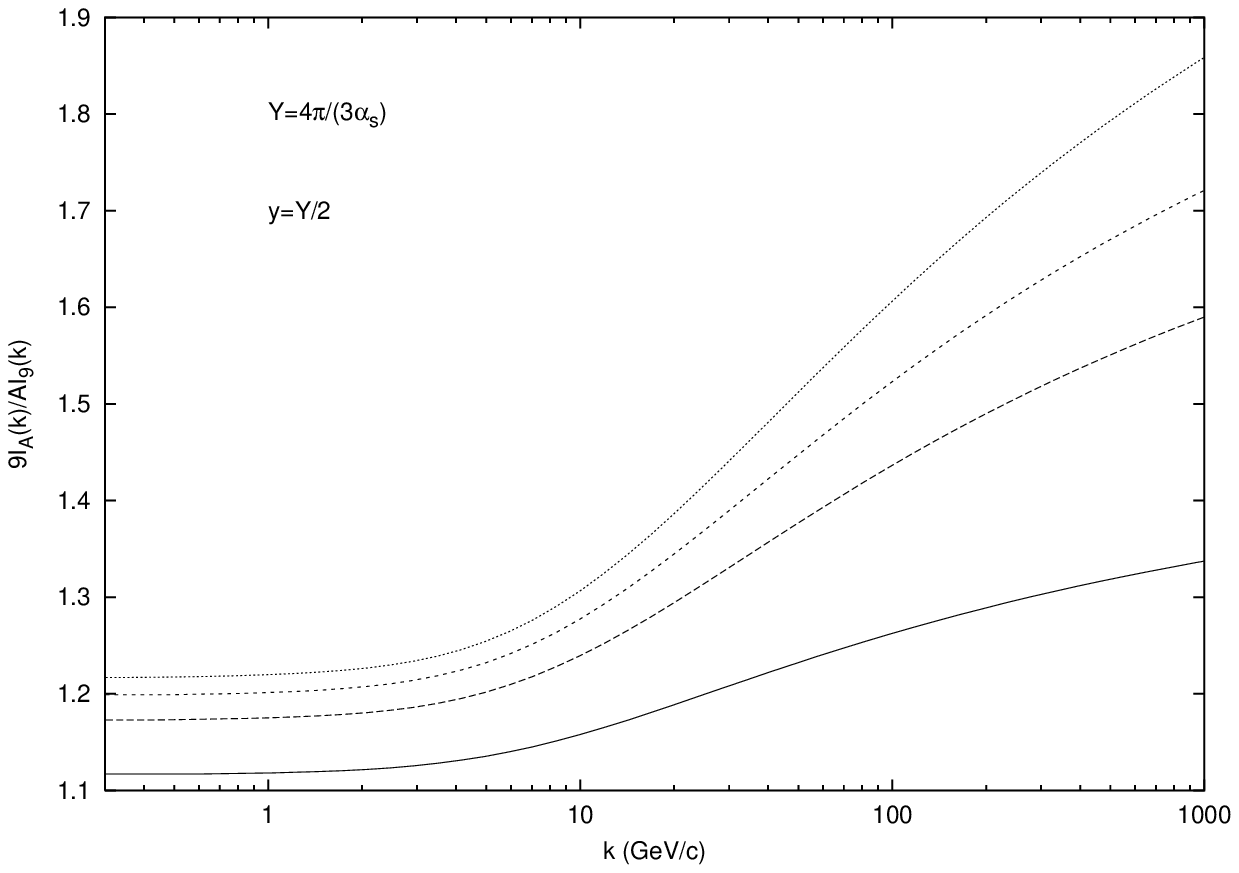}}
\caption{The ratios $(9/A)I_A(k)/I_9(k)$ for AA collisions
at $\bar{Y}=4$ and $y=Y/2$.
Curves from  bottom to top correspond to
$A=27,64,108$ and 180}
\label{Fig13}
\end{figure}
As in the pA case, we observe scaling at momenta below $Q_s$,
which tells that at such $k$ the cross-sections are $\propto A$ with
a good precision. At momenta higher than $Q_s$ they grow with $A$
faster. In our calculations we found that at very high momenta
the cross-section grow as $\propto A^{1.1}$, that is much
slowlier than $A^{4/3}$ expected from probabilistic considerations.

Multiplicities $\mu(y)$ are obtained in AA case from the integration
of $I_A(y,k)$ over $k\geq k_{min}=2$ GeV/c. In Fig. 14 we show them
divided by $A$ for different $y$ at the overall rapidity $\bar{Y}=4$.
We see approximate scaling. However the multiplicities in fact
grow somewhat faster than $A$, which is explained by the contribution
of the high momentum tail of the spectra. The change of the form of the
$y$-dependence with the overall energy is illustrated in Fig.15 where we
plot $\mu(y)/\mu(Y/2)$ at $\bar{Y}=2,4$ and 8 for $A=180$. One observes
that the peak of the multiplicity at mid-rapidity becomes narrower with the
growth of energy.
\begin{figure}[ht]
\epsfxsize 4in
\centerline{\epsfbox{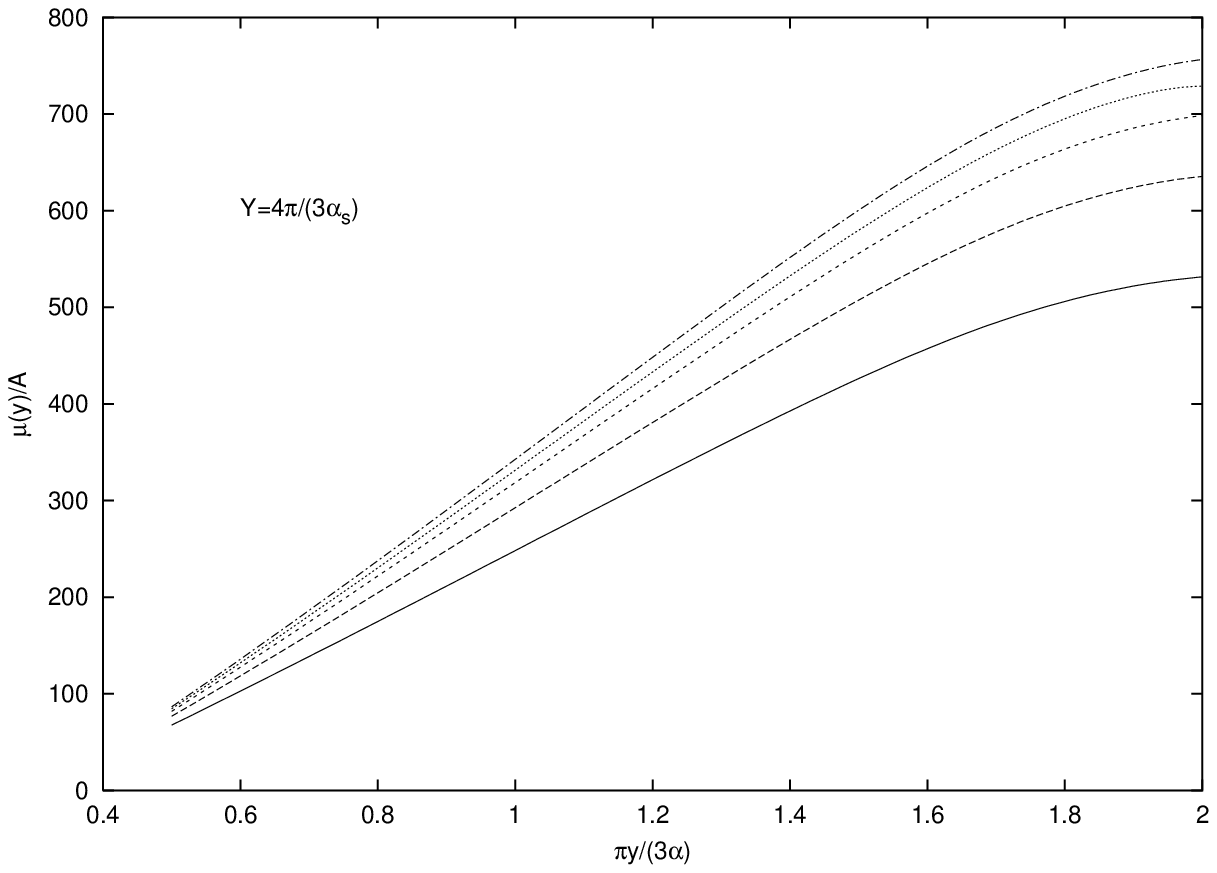}}
\caption{AA multiplicities divided by $A$ at different $y$ for
the overall rapidity $\bar{Y}=4$.
Curves from  bottom to top correspond to
$A=9,27,64,108$ and 180.}
\label{Fig14}
\end{figure}
\begin{figure}[ht]
\epsfxsize 4in
\centerline{\epsfbox{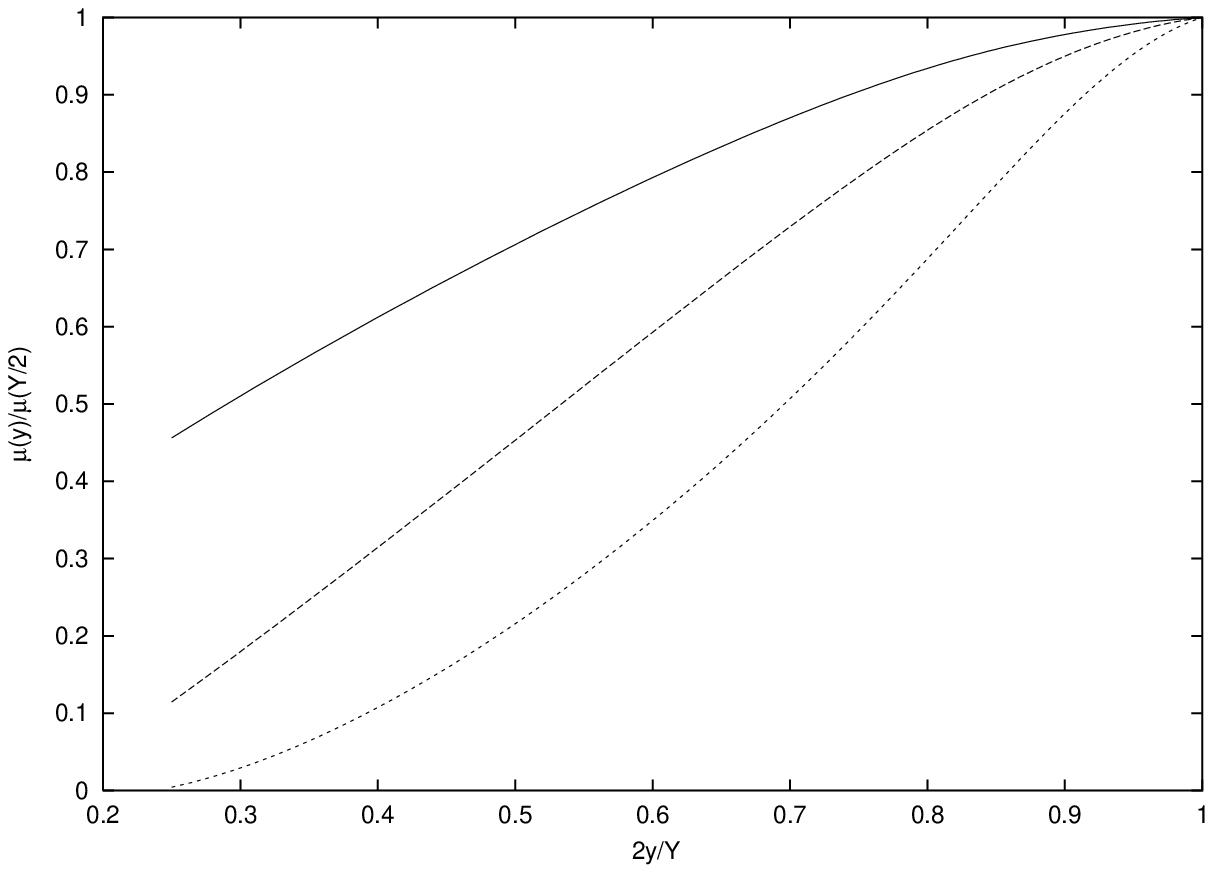}}
\caption{The form of AA multiplicities  at different energies
for $A=180$. Curves from top to bottom correspond to
$\bar{Y}=2,4$ and 8.}
\label{Fig15}
\end{figure}

All cross-sections discussed so far were obtained from the expression
(15) derived from the AGK rules. We finlaly discuss
the AA cross-sections described by the KT formula (19). As for the pA case
we present ratios of the latter cross-sections to the AGK cross-sections
in Fig. 16.
\begin{figure}[ht]
\epsfxsize 4in
\centerline{\epsfbox{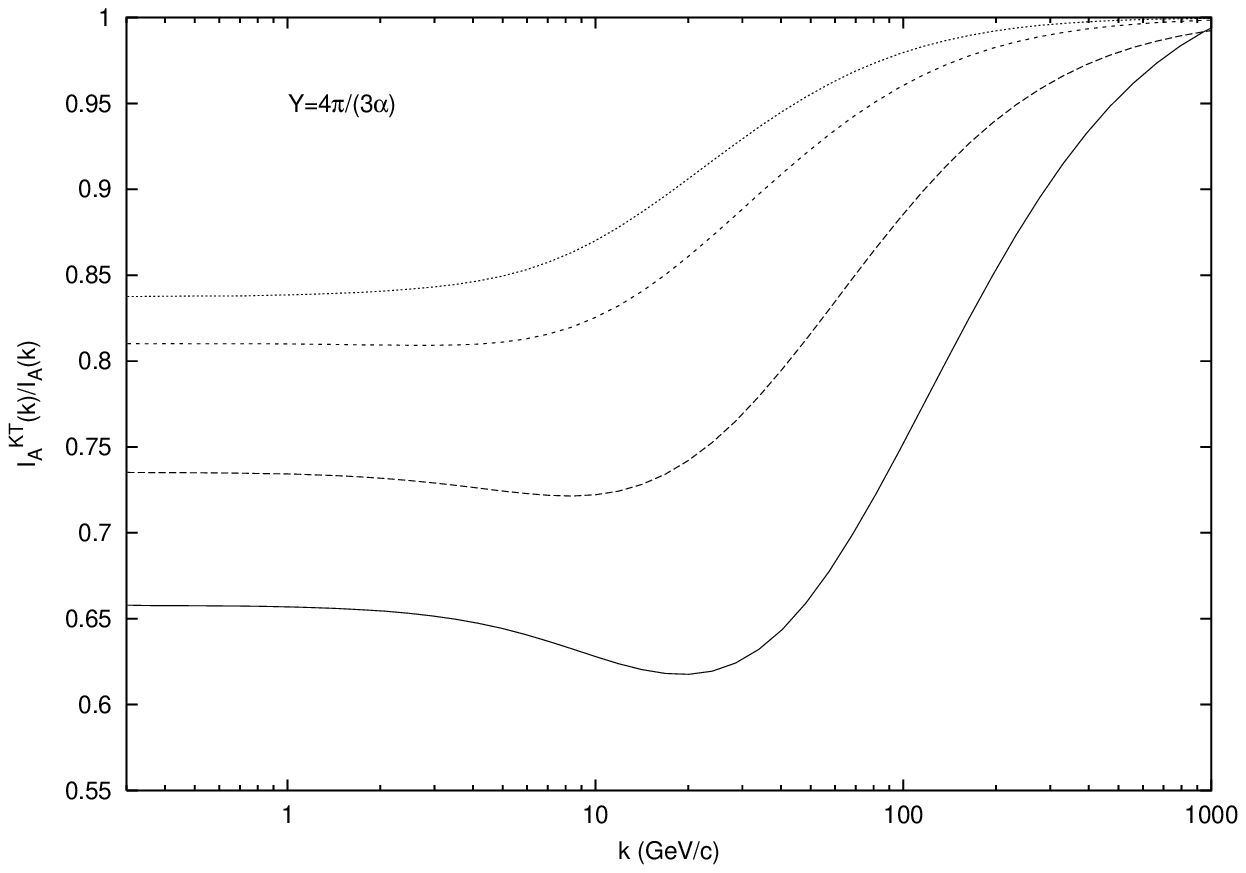}}
\caption{Ratios of the KT inclusive AA cross-sections, Eq. ( 19 ), to the
ones found from the AGK rules, Eq. ( 15 ) at $\bar{Y}=4$ for $A=180$.
Curves from bottom to top refer to $y=(1/8)Y,(1/4)Y,(3/8)Y$ and
$(1/2)Y$ (mid-rapidity) }
\label{Fig16}
\end{figure}

These ratios do not practically depend on $A$, so we show them for
$A=180$ and different $y$ at the total energy corresponding to
$\bar{Y}=4$. Their behaviour  is quite similar to the pA case.
At momenta below $Q_s$ the KT cross-sections are somewhat smaller than the
AGK ones. Their ratios grow with $y$ from $\sim 0.6$ at $\bar{y}=0.5$
to $\sim 0.8$ at $\bar{y}=2$ (mid-rapidity). Note that at $\bar{y}=0.5$
the lower part of the gluon distribution is still quite far from its
asymptotic (scaling) form. So the values for the cross-section at this
rapidity preserve a good deal of their dependence on the
initial conditions and are not characteristic for the fully evolved
dynamics of nuclear interaction. At momenta much higher than $Q_s$
all ratios tend to unity, so that the KT recipe gives the same as
the AGK one.

\section{Conclusions}
We have calculated the inclusive cross-sections and multiplicities
for gluon production in proton-nucleus collisions and
nucleus-nucleus central collisions in the
perturbative QCD hard pomeron approach with a large number of colors.
Realistic
nuclear densities were employed to account for the peripheral parts
of the nuclei, whose contribution rapidly grows with energy due to
smallness of unitarizing non-linear effects.
In contrast to the structure functions, hadronic processes required
introduction of a new parameter in the model, the value of the
strong coupling cosntant at very small energies corresponding to
the proton structure.

 The form of
the cross-sections is found to be determined by the value of the
saturation momentum $Q_s$, which depends on the rapidity and nuclear
density. At momenta much lower than $Q_s$ the spectrum is
proportional to $1/k^2$. Its $A$ dependence is close to $A^{0.7}$
for pA collisions and linear for AA collisions at $b=0$. At
momenta much higher than $Q_s$ the spectrum is found to fall
approximately as $1/k^{2.7\div 3.3}$ with the $A$-dependence as
$\sim A^{0.9}$ for pA and $\sim A^{1.1}$ for AA collisions.
The multiplicities  are found to be proportional to $A^{0.7}$ for
pA and  $A$ for AA collisions. Their peak at mid-rapidity for
AA collisions becomes more pronounced with the growth of energy.

We also compared two different forms for the inclusive
cross-section, which follow from the AGK rules or the dipole
picture. The difference between their predictions was found to be
absent for values of momenta larger than $Q_s$. At momenta smaller
than $Q_s$ the difference reduces to a universal constant factor:
the dipole cross-sections are just $\sim 0.7\div 0.8$ of the
AGK cross-sections.
With the growth of energy  this factor slowly grows towards unity,
so that it is not excluded that at infinite $Y$ the two cross-sections
totally coincide at all values of momenta.
All our conclusions about the energy, momentum and $A$ dependence
are equally valid for both forms of the inclusive cross-sections.

As mentioned in the Introduction a few more phenomenological studies
of the gluon production in nucleus-nucleus collisions were recently
made in the classical approximation to the colour-glass condensate
model ~\cite{KNV} and in the saturation model of ~\cite{KLN}. In
both studies quantum evolution was neglected, so that scaling with
the saturation momentum $Q_s$ was postulated rather than derived.
The saturation momentum thus appeared as an external parameter,
whose $A$ and $Y$ dependence were chosen on general grounds and
whose values were fitted to the experimental data at RHIC. In both
models the multiplicities turned out to be proportional to the
number of participants (modulo logarithmic dependence on $A$).
This agrees with our results.
However the form of the inclusive distributions in momenta found in
~\cite{KNV} is  different from ours. Its behaviour both at small $k$
($\sim 1/\sqrt{k^2+m^2}$ with $m=0.0358 Q_s$) and at large
 $k$ ($\sim 1/k^4 $)
disagrees with the form of the spectrum we have found. For realistic
nuclei the spectrum
was calculated in ~\cite{KNV} only up to 6$\div$7 GeV/c, so it is not
possible to see if any change in its $A$-behaviour will occur at
higher momenta.  The value of the saturation
momentum and the speed of its growth with rapidity which we have
found from the QCD pomeron model with full quantum evolution are
larger than the fitted values in both ~\cite{KNV} and
~\cite{KLN}. This is no wonder in view of a very high value of the
BFKL intercept in the leading approximation which is obtained
with the  value for the strong coupling constant at present energies.
From the
phenomenological point of view this is the main drawback of the BFKL
theory. To cure it one possibly has to include  higher orders
of the perturbation expansion and  the running coupling constant.
Although some work in this direction has been done for linear
evolution ~\cite{LFC},  no attempts to generalize this to non-linear
evolution in some rigour has been made yet. As it stands, the model
can pretend to describe the data at energies considerably above
the presently achieved.  One may hope that future  data from
experiments at LHC will be more suitable for the theoretical analysis
in the framework of the model.


\section{Acknowledgements}
The author is deeply indebted to Yu.Kovchegov for a constructive
discussion and valuable comments and to N.Armesto and B.Vlahovic for their
interest in this work. This work has been supported by
a NATO Grant PST.CLG.980287.


\begin{thebibliography}{100}
%
\bibitem{BFKL}
E.A.Kuraev, L.N.Lipatov and V.S.Fadin, Sov. Phys. JETP {\bf 45} (1977) 199;
Ya.Ya.Balitsky and L.N.Lipatov, Sov. J.Nucl.Phys. {\bf 28} (1978) 22.
%
\bibitem{kov}
Yu.V.Kovchegov, Phys. Rev. {\bf D 60} (1999) 034008; {\bf D 61}
(2000) 074018.
%
\bibitem{bra1}
M.A.Braun, Eur. Phys. J. {\bf C 16} (2000) 337.
%
\bibitem{bra2}
M.A.Braun, Phys. Lett. {\bf B 483} (2000) 115.
%
\bibitem{bra3}
M.A.Braun, Eur. Phys. J. {\bf C 33} (2004) 113.
%
\bibitem{bra4}
M.A.Braun, Phys. Lett. {\bf B 483} (2000) 105.
%
\bibitem {AGK}
 V.A.Abramovsky, V.N.Gribov and O.V.Kancheli,
Sov. J. Nucl. Phys. {\bf 18} (1974) 308
%
\bibitem{nestor}
 J.L.albacete, N.Armesto, A.Kovner, C.A.salgado and U.A.Wiedemann,
Phys. Rev. Lett {\bf 92} (2004) 082001
%
\bibitem{kov2}
D.Kharzeev, Yu.Kovchegov and K.Tuchin, hep-ph/0405045
%
\bibitem{KNV}
A.Krasnitz, Y.Nara and R.Venugopalan, Phys. Rev. Lett {\bf 87} (2001) 192302;
Nucl. Phys. {\bf A 717} (2003) 268.
%
\bibitem{KLN}
D.Kharzeev and M.Nardi, Phys.Lett. {\bf B 507} (2001) 121; D.Kharzeev and E.Levin,
Phys. Lett. {\bf B 523} (2001) 79.
%
\bibitem{KT}
 Yu. Kovchegov and K.Tuchin,  Phys. Rev. {D 65} (2002) 074026.
%
\bibitem{ML}
L.D.McLerran and R.Venugopalan, Phys. Rev. {\bf D 49} (1994) 2233, 3352;
{\bf D 50} (1994) 2225; E.G.Ferreiro, E.Iancu, A.Leonidov and L.D.McLerran,
Nucl. Phys. {\bf A 703} (2002) 489.
%
\bibitem{bra5} M.A.Braun, hep-ph/0407346, to be published in Phys. Lett. B
%
\bibitem{gobi}
K.Golec-Biernat and M.Wuesthoff, Phys. Rev. {\bf D 59} (1999) 014017;
{\bf D 60} (2000) 114023.
%
\bibitem{cia}
M.Ciafaloni {\it et al.}, Nucl. Phys. {\bf B 98} (1975) 493.
%
%
\bibitem{BW}
J.Bartels and M.Wuesthoff, Z.Physik, {\bf C 66} (1995) 157.
%
%
\bibitem{arbra} N.Armesto and M.A.Braun, Eur. Phys. J. {\bf C 20} (2001) 517.
%
\bibitem{LFC}
V.S.Fadin and L.N.Lipatov, Phys. Lett. {\bf B 429} (1998) 127;
G.Camici and M.Ciafaloni, Phys. Lett. {\bf B 412} (1998) 396;
{\bf B 430} (1998) 349.
%
\end{thebibliography}
\end{document}